\documentclass{INTERSPEECH2023}

\usepackage{graphicx}
\usepackage{arydshln}
\usepackage{subcaption}
\usepackage{algorithm}
\usepackage{algpseudocode}
\usepackage{tabularx}
\usepackage{booktabs}
\usepackage{makecell} 
\usepackage{cite}
\usepackage[utf8]{inputenc}
\usepackage{xcolor}
\usepackage{hyperref} 
\hypersetup{
    colorlinks=true,                
    breaklinks=true,                
    urlcolor= blue,
    linkcolor= black,                
    bookmarksopen=false,
    filecolor=black,
    citecolor=blue,
    linkbordercolor=blue
}
\AtBeginDocument{\hypersetup{pdfborder={0 0 0}}}

\usepackage{amsmath,graphicx}
\usepackage{url}

\makeatletter

\renewcommand{\section}{\@startsection
  {section}%
  {1}%
  {}%
  {0.2\baselineskip}%
  {0.1\baselineskip}%
  {}
}%

\renewcommand{\subsection}{\@startsection
  {subsection}%
  {2}%
  {}%
  {0.15\baselineskip}%
  {0.11\baselineskip}%
  {}}%

\renewcommand{\subsubsection}{\@startsection
  {subsubsection}%
  {3}%
  {}%
  {0.1\baselineskip}%
  {0.1\baselineskip}%
  {}}%
\makeatother

\renewcommand{\footnotesize}{\fontsize{8pt}{9pt}\selectfont}

\title{Property-Aware Multi-Speaker Data Simulation: A Probabilistic Modelling Technique for Synthetic Data Generation}
\name{Tae Jin Park, He Huang, Coleman Hooper, Nithin Koluguri,\\ Kunal Dhawan, Ante Juki\'{c}, Jagadeesh Balam and Boris Ginsburg}
\address{
  NVIDIA, Santa Clara, USA
  }
\email{\{taejinp,heh,chooper,nkoluguri,kdhawan,ajukic,jbalam,bginsburg\}@nvidia.com}
\vspace{-20px}
\begin{document}

\maketitle
 \vspace{-30px}
\begin{abstract}
We introduce a sophisticated multi-speaker speech data simulator, specifically engineered to generate multi-speaker speech recordings. A notable feature of this simulator is its capacity to modulate the distribution of silence and overlap via the adjustment of statistical parameters. This capability offers a tailored training environment for developing neural models suited for speaker diarization and voice activity detection. The acquisition of substantial datasets for speaker diarization often presents a significant challenge, particularly in multi-speaker scenarios. Furthermore, the precise time stamp annotation of speech data is a critical factor for training both speaker diarization and voice activity detection. Our proposed multi-speaker simulator tackles these problems by generating large-scale audio mixtures that maintain statistical properties closely aligned with the input parameters. We demonstrate that the proposed multi-speaker simulator generates audio mixtures with statistical properties that closely align with the input parameters derived from real-world statistics. Additionally, we present the effectiveness of speaker diarization and voice activity detection models, which have been trained exclusively on the generated simulated datasets.
\end{abstract}

\section{Introduction}
The evolution of deep neural network models within the realm of speech signal processing has significantly enhanced the performance and precision of the machine learning systems \cite{park2022review}. These advances have facilitated an end-to-end training approach, allowing the entire model to be optimized, transforming raw audio input into meaningful labels. However, achieving competitive accuracies with these neural models depends on the procurement of a substantial amount of data. This data is integral to ensuring generalizability and improving accuracy.

Obtaining sufficient training data in certain domains poses a significant challenge due to an array of factors. In speech signal processing field, the challenges are concentrated on privacy concerns, data-imbalance issues, limited availability and the financial cost of data collection. The task becomes even more demanding when it involves speaker diarization. This increased difficulty is primarily because speaker diarization requires a complex dataset with multiple speakers, embodying a broad range of variabilities. These variabilities encompass aspects such as gender, acoustic conditions, and conversation types. Hence, the development and optimization of effective deep neural network models for speech signal processing, particularly speaker diarization, hinges on overcoming these challenges related to data collection.

In response to the challenge of data scarcity in specific fields, the machine learning community has adopted synthetic data, which mitigates the aforementioned issues to a certain degree. In order for synthetic data to be effective, the generated data should capture the characteristics and patterns of real-world data (\textit{realism}) while maintaining a broad range of variations (\textit{diversity}). Also, accurate and consistent labeling of the synthetic dataset is essential. Additionally, in speaker diarization or Voice Activity Detection (VAD), the diversity of speakers, sentence length, and frequency of speaker turns in conversations should be well balanced, mirroring real-world data.
\begin{figure}[t]
\centering
\includegraphics[width=0.95\columnwidth]{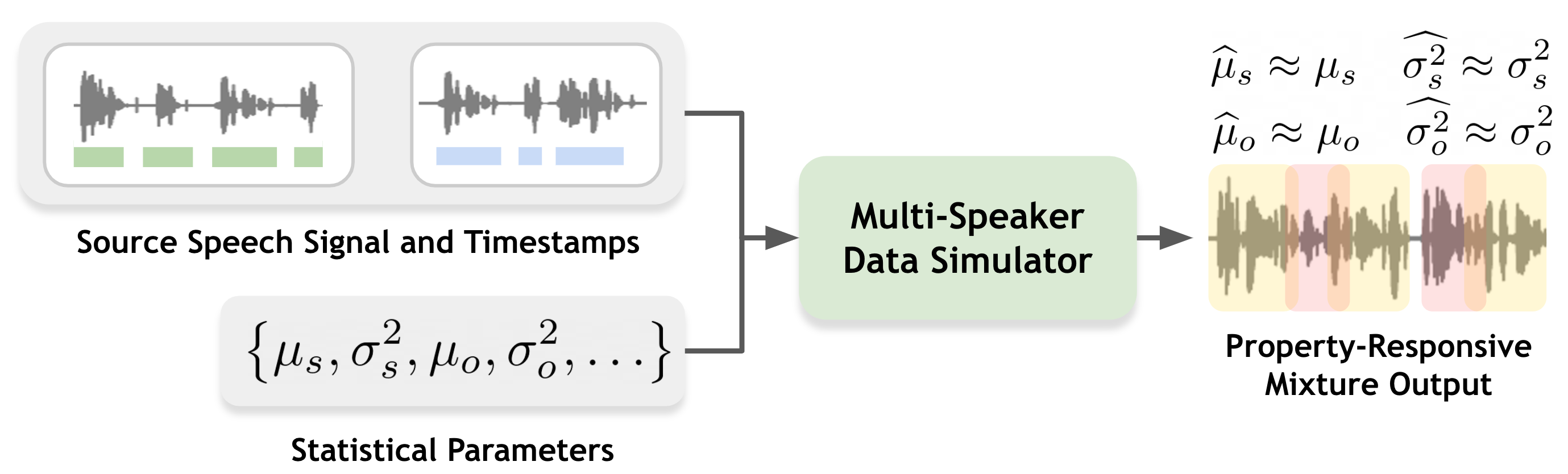}
\vspace{-8px}
\caption{Property-aware multispeaker data simulator that generates targeted amount of pause and overlap.}
\label{fig:intro}
\vspace{-16px}
\end{figure}

Until now, in the fields of speech recognition and speaker diarization, most published articles have focused on data augmentation techniques, such as the widely used SpecAugment \cite{park2019specaugment} or the data augmentation speech recognition toolkit \cite{ko2015audio, povey2011kaldi}. There exist simulation tools (e.g., one featured in \cite{fujita2019end}) initially developed for source separation but often utilized in training speaker diarization systems \cite{fujita2019end,hershey2016deep,horiguchi2020end}. Recently, a multi-speaker data simulator for speaker end-to-end speaker diarization also appeared in \cite{landini2022simulated,landini2023multi}, which tries to create mixtures that resemble the pauses and overlaps of the real-world audio recordings. While the data simulation techniques introduced in \cite{hershey2016deep, landini2023multi} serve their purpose very well, these data simulation techniques tend to employ a range of parameters which do not explicitly correlate with specific properties such as pauses and overlaps within the resulting simulated speech recordings. Consequently, even though the previously proposed simulation systems accept numerous parameters, their lack of control over the generated signal could lead to unpredictability in the amount of silence and overlap.

In this work, we introduce a dynamic sampling technique that constantly reflects the discrepancy between the generated data and the targeted amount of overlap speech and silence employing probabilistic models for precision and control. We refer to such feature as \textit{``Property-aware simulation''}. As illustrated in Fig.~\ref{fig:intro}, the proposed multi-speaker data simulator takes speech signal and its alignment (time stamps) and blends these signals to simulate multi-speaker audio recordings. Herein, we elaborate on the guiding principles for our data simulation systems:

\begin{itemize}
\item The simulated sessions are designed to incorporate the required amount of silence, overlap, and sentence length based on statistical analysis.
\item The speech signal generated by the simulation system exhibits a significant level of variability across sessions, including overlap ratio, silence ratio, and average sentence length.
\item The simulation system employs parallel processing techniques, leveraging multiple graphics processing units (GPUs), enabling large-scale data generation at higher speed.
\item The implementation of the data simulator is open-source and publicly available online.\footnote{\url{https://github.com/NVIDIA/NeMo/main/tools/speech_data_simulator}}

\end{itemize}
\section{System description} 
\subsection{Major parameters}

\begin{figure}
    \centering
    \includegraphics[scale=0.45]{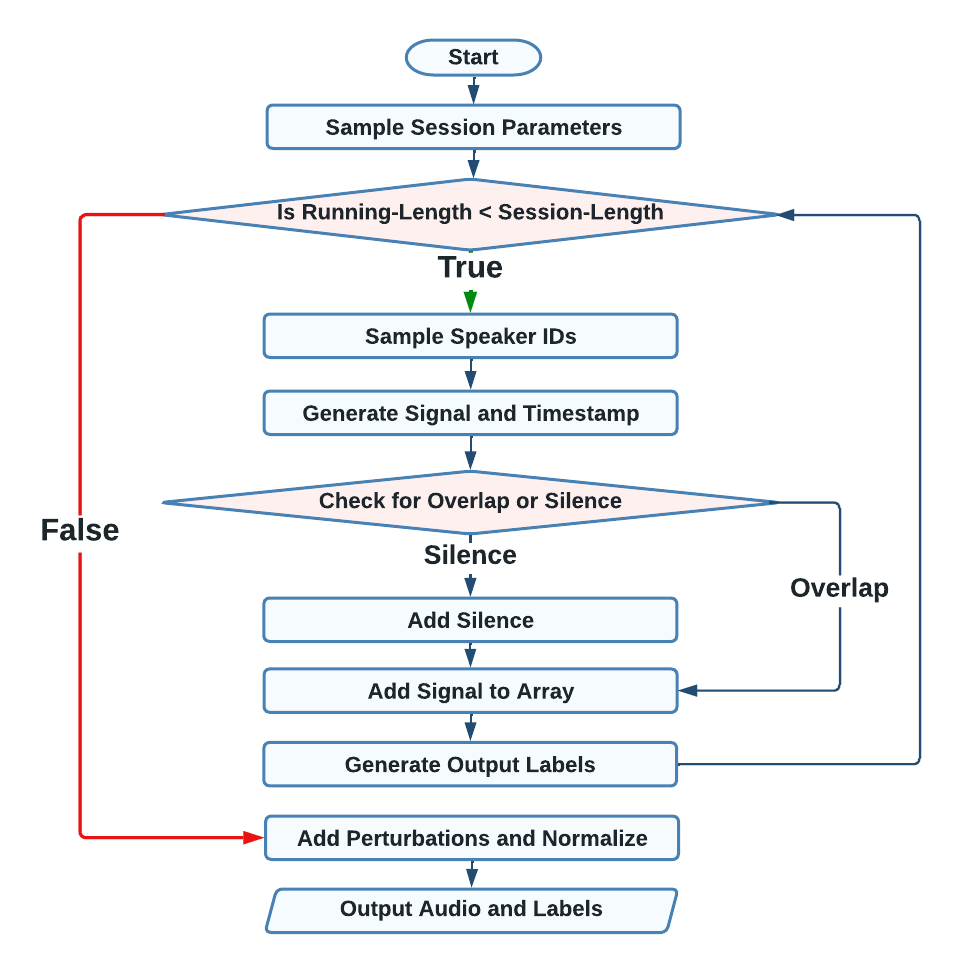}
    \vspace{-15pt} 
    \caption{Flowchart of the proposed multi-speaker data simulator}
    \label{fig:flowchart}
    \vspace{-12pt} 
\end{figure}

A flow diagram of the proposed system is shown in Fig.~\ref{fig:flowchart}. In the following sections we describe the main parameters and implementation details. Note that the following parameters are the most crucial subset of parameters that are determined before starting data simulation:
\begin{itemize}
    \item Session length $L_{S}$: A floating point number that determines the total duration of the created session in second.
    \item Number of sessions $N_{S}$: An integer to determine the number of session to be simulated, so that the total duration of the generated data is $L_{S}\cdot N_{S}$ seconds.
    \item Number of speakers $N_{spk}$: An integer number that determines how many speakers in a session.
    \item Turn Probability $p_{\text{turn}}$: A floating point number that determines the speaker turn change from one to another.
    \item Overlap ratio mean $\mu_o$ and variance $\sigma_o^2$: Parameters that determine the distribution of overlap. 
    \item Silence ratio mean $\mu_s$ and variance $\sigma_s^2$: Parameters that determine the distribution of silence. 
\end{itemize}

The following are random variables that are created at each session:
\begin{itemize}
    \item Sentence length $s_l$ determines how many words are included in a newly added utterance (also referred to as a sentence).
    \item Silence length $\widetilde{m}_{s}$ determines the duration between sentences.
    \item Overlap length $\widetilde{m}_{o}$ determines how much portion of speech is overlapped with the following speech segment.
\end{itemize}

\subsection{Session Parameter Sampling}
The following variables are sampled during the very first step named ``\textit{Sample Session Parameters}'' in Fig.~\ref{fig:flowchart}.

\begin{itemize}
\item Session random seed sampling: 
Set a random seed value which would be used to create a reproducible simulation environment. 
\item Set speaker dominance: Call a method that randomly determines the dominance of each speaker in the session.
\item Speaker volumes: Set the volume level of each speaker in the session.

\item Setting Session Silence and Overlap Mean: To control the amount of silence and overlap in a session, we can set the mean values for these parameters using the following equation, which describes the method of moment estimates \cite{fielitz1975concepts} for a Beta distribution\footnote{Beta distribution is employed due to its compatibility with the range of overlap and silence ratios, which fall within its support of [0, 1], and its capacity to model skewed distributions\cite{heldner2010pauses}.} :
\begin{align}
\vspace{-5px}
\alpha_{{\{o, s\}}} & = \frac{{\mu_{{\{o, s\}}}^2 \cdot (1 - \mu_{{\{o, s\}}})}}{{\sigma_{{\{o, s\}}}^2}} - \mu_{{\{o, s\}}} \\
\beta_{{\{o, s\}}} & = \frac{{\mu_{{\{o, s\}}} \cdot (1 - \mu_{{\{o, s\}}})^2}}{{\sigma_{{\{o, s\}}}^2}} - (1 - \mu_{{\{o, s\}}}) 
\end{align}
Here, $\mu$ represents the mean ratio of silence or overlap and $\sigma$ represents its variance. These parameters are fed by the user to control the overall silence and overlap ratio. To ensure that $\alpha_{{\{o, s\}}}\text{$>$}0$ and $\beta_{{\{o, s\}}}\text{$>$}0$, the input mean and variance values should be within the following range:
\begin{equation}
\begin{cases}
0 < \mu_{{\{o, s\}}} < 1 \\
0 < \sigma^2_{{\{o, s\}}} \leq \mu_{{\{o, s\}}} (1 - \mu_{{\{o, s\}}})
\end{cases}
\end{equation}
We can then sample the session silence mean $X_{\mu_{s}}$ and session overlap mean $X_{\mu_{o}}$ from the Beta distribution, as follows:
\begin{align}
X_{\mu_{s}} & \sim Beta(\alpha_{s}, \beta_{s}), \\
X_{\mu_{o}} & \sim Beta(\alpha_{o}, \beta_{o}).
\end{align}
Here, $\alpha_{s}$ and $\beta_{s}$ are based on the mean $\mu_{s}$ and variance $\sigma_{s}$ for the silence ratio in a session, while $\alpha_{o}$ and $\beta_{o}$ are based on the mean $\mu_{o}$ and variance $\sigma_{o}$ for the overlap speech ratio in a session. By setting the session silence and overlap mean values in this way, we can control the amount of silence and overlap in a session, which follows Beta distribution.

\end{itemize}

\subsection{Sampling Routine for Data simulation}
The following provides a description of each step involved in generating a simulated multi-speaker audio recording. In this section: $n_s$ denotes the current sample count, $s_{spk}$ the speaker index, and $\Tilde{L}_{S}$ the running length of the audio signal thus far.

\subsubsection{Data synthesis loop}

As described in the Algorithm \ref{alg:cap}, the running length of the current session $\Tilde{L}_{S}$ is monitored at every loop and while the condition $\Tilde{L}_{S} < L_{S}$ is held, the sampling process is continued until the running length $\Tilde{L}_{S}$ exceeds the desired length $L_{S}$.

\subsubsection{Sample Speaker ID}
The turn probability $p_{\text{turn}}$ is compared with a value drawn from a uniform distribution. 
\begin{align}
U(0,1) < p_{\text{turn}}
\end{align}
If the sampled value is less than the $ p_{\text{turn}}$ value, a randomly chosen speaker is selected from the pre-determined speaker group, for example $\mathcal{S}_{spks}=\{s_1, s_2, \dots, s_{N_{spk}} \}$.

\subsubsection{Build Sentence}
The parameter $s_{l}$, which represents sentence length, is assumed to follow a negative binomial distribution. This approach is based on the probabilistic model for word-level sentence length (measured in words) of human language as detailed in \cite{jin2017will}.
\begin{align}
s_{l} & \sim {\text{NB}}(k_{w}, p_{w}) \\ 
P_{\text{NB}}(X = k_{w}) &= \binom{X+k_w-1}{k_{w}-1} p_w^{k_{w}} (1-p_w)^X
\end{align}
Based on the sentence length $s_l$ and speaker  (also referred as speaker turn) $s_{\text{spk}}$, we randomly select the given number of words from the forced-alignment data. This process is denoted as $\Call{BuildSentence}$ function in the Algorithm \ref{alg:cap}.

\begin{equation}    
\label{eq:add_sent}
\Tilde{L}_{spch}, \Tilde{L}_{sil} = \Call{BuildSentence}{s_{l}, s_{spk}}.
\end{equation}

\subsubsection{Overlap-Silence Selector}
In this step, the data-simulator system compares the current silence ratio to the current overlap ratio. Thus, at each utterance loop, it switches to either silence or overlap mode according to the amount of the gap between current ratio and session mean in configurations.
\begin{align}
\Delta S = \frac{\Tilde{L}_{sil}}{\Tilde{L}_{S}} - \mu_{s}\\
\Delta O = \frac{\Tilde{O}_{spch}}{\Tilde{L}_{spch}} - \mu_{o}
\end{align}
We employ two different quantities: \textit{silence discrepancy} $\Delta S$ represents  the gap between desired silence time and the current silence time and \textit{overlap discrepancy} $\Delta O$ represents  which is the gap between desired overlap speech and the current overlap speech time. We choose whichever is smaller than other.

\subsubsection{Estimating the Required Overlap Amount}
Overlap $\widetilde{m}_{o}$ is calculated so that the newly added amount of overlap matches the expected amount of $X_{\mu_o}$.
\begin{equation}
X_{\mu_{o}} = \frac{\widetilde{m}_{o} + \Tilde{O}_{spch}}{\Tilde{L}_{spch} - \widetilde{m}_{o}}
\end{equation}
Afterwards, we solve for $\widetilde{m}_{o}$ and assign it the value derived from the following equation:
\begin{equation}
\widetilde{m}_{o} \gets \frac{X_{\mu_{o}}\Tilde{L}_{spch} - \Tilde{O}_{spch}}{X_{\mu_{o}} + 1}
\end{equation}

\subsubsection{Estimating the Required Silence Amount}
We set up an equation that matches the expected amount of silence after adding the silence (denoted by $\widetilde{m}_{s}$) with the sampled mean $X_{\mu_{s}}$  as follows:
\begin{equation}
X_{\mu_{s}} = \frac{\widetilde{m}_{s} + \Tilde{L}_{sil}}{\widetilde{m}_{s} + \Tilde{L}_{S}}
\end{equation}
Solve for $\widetilde{m}_{s}$ then we assign the following value:
\begin{equation}
\widetilde{m}_{s} \gets \frac{\Tilde{L}_{sil} - X_{\mu_{s}}\Tilde{L}_{S}}{X_{\mu_{s}} - 1}
\end{equation}

\subsubsection{Sampling overlap and silence amount}
For both silence and overlap cases, we employ gamma distribution since gamma distribution is continuous version of negative binomial distribution that is used to model sentence length in \cite{jin2017will} Thus, we model the distribution of the two continuous quantity, silence and overlap length, as following equations:
\begin{align}
k \gets \widetilde{m}^{2} / \sigma^{2} \\
\theta \gets \sigma^{2} / \widetilde{m} \\
x_{\Delta t} \sim \Gamma(k, \theta),
\end{align}
where $x_{\Delta t}$ is the sampled silence amount $s_{\Delta t}$ in silence case and overlap amount $o_{\Delta t}$ in overlap case.

\begin{algorithm}[t]
\caption{Dialogue Simulation}\label{alg:cap}
\begin{algorithmic}
\Require 
$L_{S}$,  $\sigma^{2}_{d}$  $\mu_o$,  $\mu_s$, $p_{\text{turn}}$ $\sigma_{o}^2$, $\sigma_{s}^2$   $\in$ $\mathbb{R}$ and $N_{spk} \in \mathbb{N}$ \\
$p \in (0, 1] \vee \mu_{d}, \mu_{o}, \mu_{s}, \sigma_{d}^2, \sigma_{o}^2, \sigma_{s}^2 \in [0, 1]$

\State $(\alpha_{s},\beta_{s}) \gets \Big(\mu_{s}^2 \frac{(1 - \mu_{s})}{\sigma^{2}_{s}} - \mu_{s}, \mu_{s}\frac{(1 - \mu_{s})^{2}}{\sigma^{2}_{s}} - ( 1 - \mu_{s} )\Big)$
\State $X_{\mu_{s}} \sim Beta(\alpha_{s}, \beta_{s})$ \Comment{Sample session silence rate mean}

\State $(\alpha_{o}, \beta_{o}) \gets \Big( \mu_{o}^2 \frac{(1 - \mu_{o})}{\sigma^{2}_{o}} - \mu_{o}, \mu_{o}\frac{(1 - \mu_{o})^{2}}{\sigma^{2}_{o}} - (1 - \mu_{o}) \Big)$
\State $X_{\mu_{o}} \sim Beta(\alpha_{o}, \beta_{o})$ \Comment{Sample session overlap rate mean}

\While{$\Tilde{L}_{S} < L_{S}$}
\If{$U(0,1) < p_{\text{turn}}$} 
\State $s_{spk}$ = $\Call{GetNextSpeaker}{\mathcal{S}_{spks}, s_{spk}}$
\EndIf
\State $s_{l} \sim NB(k_{w}, p_{w})$
\State $\Tilde{L}_{spch}$, $\Tilde{L}_{sil}$ $\gets$ \Call{BuildSentence}{$s_{l}$, $s_{spk}$}
\State $\Delta S \gets \frac{\Tilde{L}_{sil}}{\Tilde{L}_{S}} - \mu_{s}$ \Comment{Silence deficiency}
\State $\Delta O \gets \frac{\Tilde{O}_{spch}}{\Tilde{L}_{spch}} - \mu_{o}$ \Comment{Overlap deficiency}
\If{$\Delta S \leq \Delta O $}
\State $\widetilde{m}_{s} \gets \frac{\Tilde{L}_{sil} - X_{\mu_{s}}\Tilde{L}_{S}}{X_{\mu_{s}} - 1}$
\State $k_{s} \gets \widetilde{m}_{s}^{2} / \sigma^{2}_{s}$ 
\State $\theta_{s} \gets \sigma^{2}_{s}/ \widetilde{m}_{s} $
\State $s_{\Delta t} \sim \Gamma(k_{s}, \theta_{s})$
\State \Call{AddSentence}{$s_{\Delta t}$, 0}
\ElsIf{$\Delta S > \Delta O $}
\State $\widetilde{m}_{o} \gets  \frac{X_{\mu_{o}}\Tilde{L}_{spch} - \Tilde{O}_{spch}}{X_{\mu_{o}} + 1} $
\State $k_{o} \gets \widetilde{m}_{o}^{2} / \sigma^{2}_{o}$ 
\State $\theta_{o} \gets \sigma^{2}_{o} / \widetilde{m}_{o}$
\State $o_{\Delta t} \sim \Gamma(k_{o}, \theta_{o})$
\State \Call{AddSentence}{0, $o_{\Delta t}$}
\EndIf
\EndWhile
\vspace{-1px}
\end{algorithmic}
\end{algorithm}

\section{Experimental Results}

\label{section:experimental results}

\subsection{Data simulation test}
In this section, we demonstrate that the proposed data simulator can simulate the multi speaker data with the given parameters. To check wether the simulator generates data which has the distribution we intended to create, we compare the simulated data with the statistics extracted from real-world datasets. The overlap mean, overlap mean variance, silence mean and silence mean var values are collected from the real-world datasets and fed to the simulator. 



\begin{table}[t]
\centering
\caption{Simulated vs. Real-world Dataset Statistics}
\vspace{-10px}
\begin{tabular}{ r| r| c c}
\Xhline{1.5pt}
\textbf{Dataset} & \textbf{Type} & \textbf{Mean} & \textbf{Var.} \\
\Xhline{1.5pt}
CH. Simul. & observed sil. ratio & 0.1409 & 0.0045 \\
CHAES & real-world sil. ratio & 0.1473 & 0.0061 \\
\hline
CH. Simul. & observed ovl. ratio & 0.0759 & 0.0019 \\
CHAES & real-world ovl. ratio & 0.0754 & 0.0020 \\
\Xhline{1.5pt}
AMI Simul. & observed sil. ratio & 0.1804 & 0.0077 \\
AMI & real-world sil. ratio & 0.1814 & 0.0081 \\
\hline
AMI. Simul. & observed ovl. ratio & 0.1711 & 0.0092 \\
AMI & real-world ovl. ratio & 0.1473 & 0.0047 \\
\Xhline{1.5pt}  
\end{tabular}
\label{table:stats}
\vspace{-12px}
\end{table}

\begin{figure}[t]
  \centering
  \begin{subfigure}[b]{0.49\columnwidth} 
    \includegraphics[width=\textwidth]{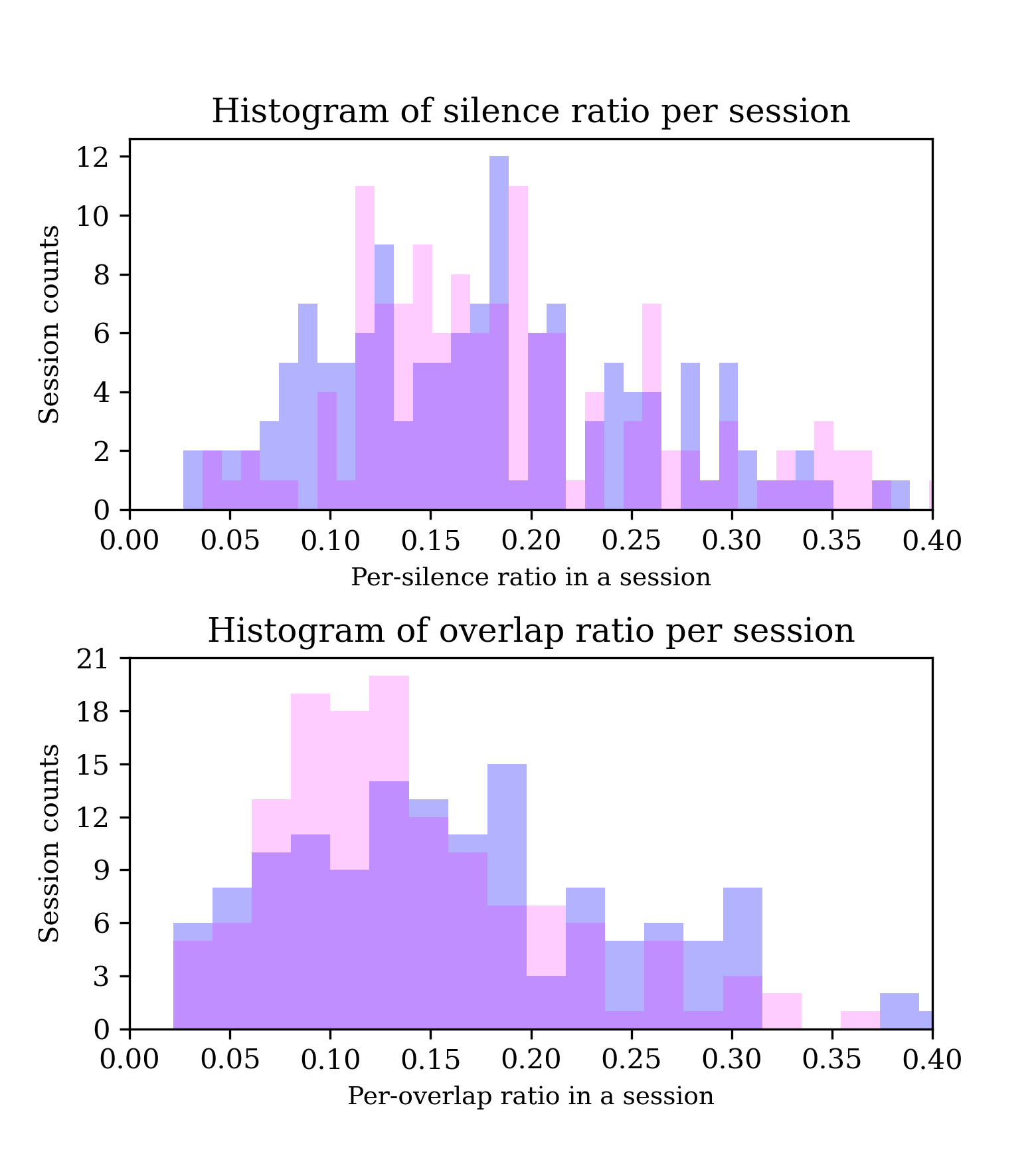}
    \vspace{-18px}
    \caption{AMI-train 139 sessions}
    \vspace{-10px}
    \label{fig:image1}
  \end{subfigure}
  \hfill 
  \begin{subfigure}[b]{0.49\columnwidth} 
    \includegraphics[width=\textwidth]{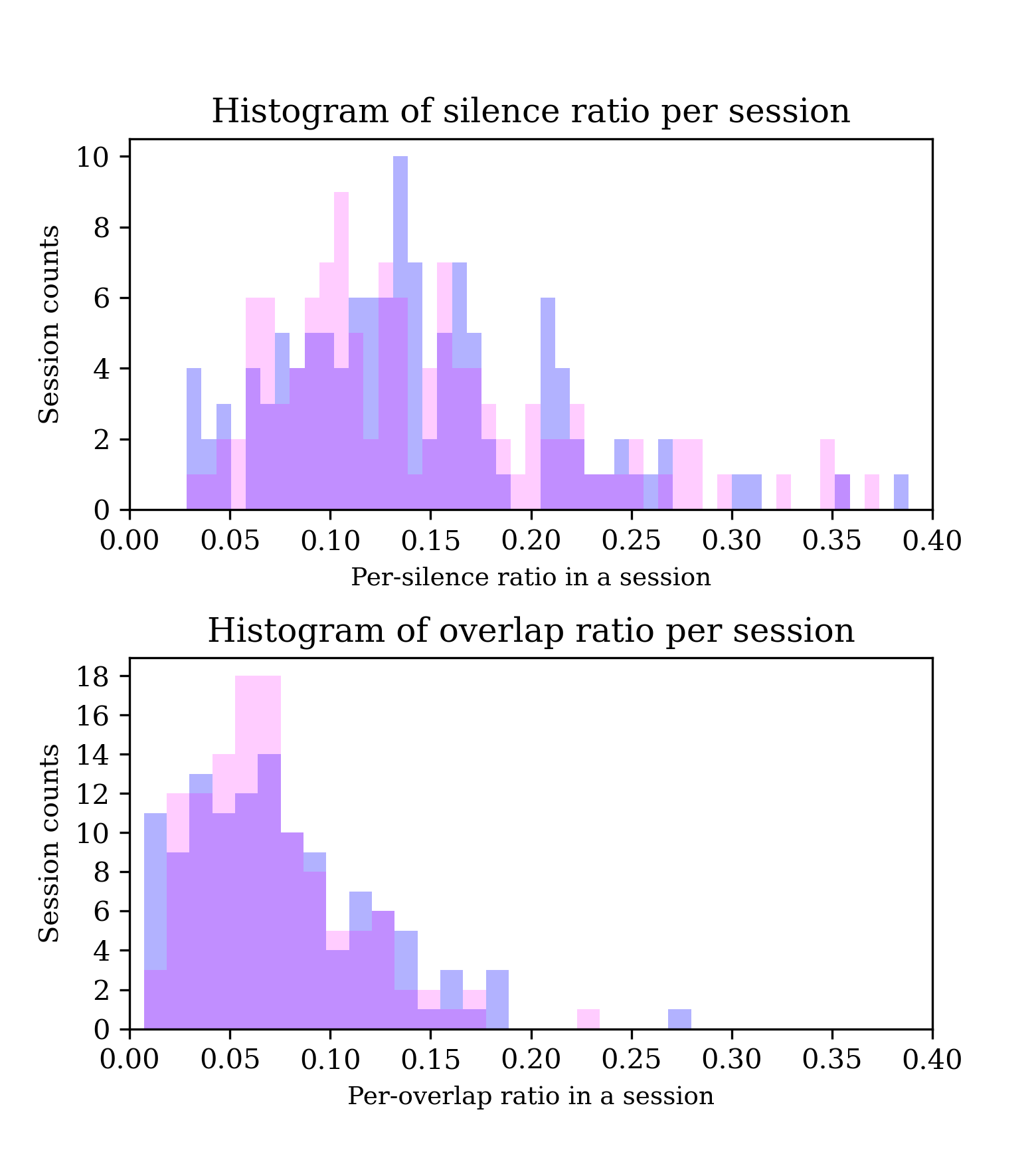}
    \vspace{-18px}
    \caption{CH109 - 109 sessions}
    \vspace{-10px}
    \label{fig:image2}
  \end{subfigure}
\caption{Histograms: Real-World (Magenta) vs. Simulated (Blue) Data; Overlaps in Purple.}
  \vspace{-22px}
  \label{fig:hist_comparison}
\end{figure}

Table \ref{table:stats} presents a quantitative comparison of observed values derived from both the simulated and real-world datasets for \textit{train} split of AMI(MixHeadSet)~\cite{kraaij2005ami} and CallHome American English Speech (CHAES)~\cite{canavan1997callhome}, highlighting key metrics such as the mean and variance of silence (sil.) and overlap (ovl.) ratios. Notwithstanding certain disparities between the statistics discerned from the simulated dataset and those from the real-world dataset, the simulation effectively echoes the trends characteristic of the original statistics. For an expanded analysis of this simulation's distribution, please refer to Fig. \ref{fig:hist_comparison}, which exhibits histograms contrasting the original and simulated datasets in terms of overlap and silence mean/variance. 

\subsection{Voice Activity Detector Model}
We trained a modified version of the Voice Activity Detection (VAD) model~\footnote{\url{https://catalog.ngc.nvidia.com/orgs/nvidia/teams/nemo/models/vad_multilingual_frame_marblenet}} proposed in \cite{jia2021marblenet}, using our simulated data. As source datasets, we employed Fisher English Corpus \cite{cieri2004fisher} and LibriSpeech Corpus\cite{panayotov2015librispeech}. For Fisher dataset, we use energy based VAD to filter out salient speech samples and randomly segmented audio in a range of [0.2, 0.8] seconds word length. For LibriSpeech, we use the forced alignment result in \cite{corentinj2023librispeechalignments}. We utilize two datasets: Dataset D1 comprises 0.5k hours of data from each of the LibriSpeech and Fisher English datasets. Dataset D2 consists of 1k hours from each of the LibriSpeech and Fisher datasets, supplemented by an additional 2.5k hours of multilingual data we have gathered from \cite{ardila2019common, sovaai_sova_dataset_2023, du2018aishell, pratap2020mls}.

The performance of this modified model across various speech datasets is outlined in Table \ref{subtable:vad trained on simulated dataset}, where area under the receiver operating characteristic (AUROC) is used as the metric. For the DIHARD3~\cite{ryant2020third} dataset, we excluded Conversational Telephonic Speech (CTS) and computed a macro-average across ten different domains, as the CTS domain is derived from the Fisher dataset, which possesses significantly loose timestamps. Through these modifications, we achieved an overall high performance with our model on the four datasets, especially with the application of noise augmentation and gain perturbation.

Several significant observations arise from our experiments with VAD models using the simulated dataset. Firstly, loose timestamps, which encapsulate non-speech signals at the start and end of each segment, can markedly degrade the performance of VAD. This issue is exacerbated by data augmentation, as the model is then trained with the added noise at the boundaries of each segment. Secondly, gain perturbation is a necessary consideration as the VAD model frequently overlooks low-volume speech signals. To mitigate this, the model should be trained with substantial variation in gain during the creation of audio mixtures. Lastly, the addition of overlapping speech is also essential for enhancing performance, as overlapping speech can lead to an increase in missed detections.

\begin{table}[t]
\caption{Evaluation of models on different parameters}
\vspace{-10px}
\centering
\setlength{\tabcolsep}{3pt}
\renewcommand{\arraystretch}{1.0}

\begin{subtable}{\linewidth}
\centering
\caption{AUROC for VAD task}
\vspace{-5px}
\begin{tabularx}{0.9\columnwidth}{r|c|c|c|l}
\toprule[1.5pt]
\textbf{Training Data} & \textbf{DH3} & \textbf{VoxConv} & \textbf{AMI} & \textbf{CH109} \\ 
\textbf{Dataset Split} & dev & dev & dev & - \\
\midrule
D1,$\mu_{s}$=0.5 & 87.71 & 96.15 & 95.7 & 88.07 \\
D1,$\mu_{s}$=0.3 & 89.83 & 96.19 & 94.69 & 91.04 \\
+ Gain. Aug. & 93.7 & 96.02 & \textbf{96.55} & 88.73 \\
+ D2 + Noise. Aug. & \textbf{93.96} & \textbf{97.42} & 96.04 & \textbf{92.43} \\
\bottomrule[1.0pt]
\end{tabularx}
\label{subtable:vad trained on simulated dataset}
\end{subtable}

\begin{subtable}{\linewidth}
\begin{center}
\centering
\caption{DER(\%) on Diarization Datasets}
\vspace{-5px}
\begin{tabularx}{0.95\columnwidth}{r|c|c|c|c}
\toprule[1.5pt]
\textbf{Training Data} & \textbf{DH3} & \textbf{VoxConv} & \textbf{AMI} & \textbf{CH-109} \\ 
\textbf{Dataset Split} & eval & test & eval & - \\
\midrule
LibriVox-3Kh $\mu_{o}$=0.07 & 14.49 & 6.01 & 15.96 & \textbf{9.94} \\
LibriVox-3Kh $\mu_{o}$=0.15 & \textbf{14.38} & \textbf{5.72} & \textbf{15.89} & 10.03 \\
\bottomrule[1.0pt]
\end{tabularx}
\label{subtable:diarizer trained on simulated dataset tested on well-known diarization datasets}
\end{center}
\vspace{-10px}
\end{subtable}

\begin{subtable}{\linewidth}
\begin{center}
\centering
\caption{DER(\%) on CHiME7 Datasets}
\vspace{-5px}
\begin{tabularx}{0.87\columnwidth}{r|c|c|c}
\toprule[1.5pt]
\textbf{Training Data} & \textbf{Chime6} & \textbf{Dipco} & \textbf{Mixer6} \\ 
\textbf{Dataset Split} & dev & dev & dev \\
\midrule
LibriVox-3Kh $\mu_{o}$=0.07 & 45.01 & 32.50 & 17.35 \\
LibriVox-3Kh $\mu_{o}$=0.15 & \textbf{44.37} & \textbf{31.07} & \textbf{17.13} \\
\bottomrule[1.0pt]
\end{tabularx}
\label{subtable:diarizer trained on simulated dataset tested on CHiME7 Challenge Datasets}
\end{center}
\end{subtable}
\vspace{-20px}
\end{table}

\subsection{Speaker Diarization Model}
As in the previous section, we train a modified speaker diarization model, based on \cite{park2022multi}, alongside the speaker embedding model from \cite{koluguri2022titanet}. The experiment utilizes 1k hours of LibriSpeech and 2k hours of VoxCeleb 1 and 2 ~\cite{nagrani2017voxceleb}. We use the same type of time stamps and random word-level alignment as in the Fisher dataset for Voice Activity Detection (VAD). For diarization evaluation, we use the VAD model from Table \ref{subtable:vad trained on simulated dataset}. Diarization error rate (DER) is calculated using a 0.25 sec collar, with overlap considered. DER is assessed on DIHARD3, VoxConverse-3~\cite{huh2023voxsrc}, AMI eval(test)-sets, and 2-speaker CHAES subset, CH109. Tables \ref{subtable:diarizer trained on simulated dataset tested on well-known diarization datasets} and \ref{subtable:diarizer trained on simulated dataset tested on CHiME7 Challenge Datasets} show performance variations with different synthetic dataset settings.
\section{Conclusions}
In this paper, we introduce a property-aware data simulator capable of reflecting statistics provided by the user or extracted from real-world data. The proposed data simulator utilizes an online sampling technique, allowing the system to generate a predetermined quantity of silence and overlap speech while adhering to the given probability distributions. Consequently, the generated dataset can be leveraged to train VAD models and speaker diarization models, providing highly accurate ground-truth timestamps, which is a critical element for both speech activity detection and speaker diarization. Potential future research could involve adapting this system for online generation, whereby users could supply a source dataset and generate the training dataset on-the-fly. We anticipate that the proposed data simulator will be adopted by the speech signal processing community for training neural models related to speech signals, which necessitate accurate ground truth timestamps and highly customizable speech training data.

\bibliographystyle{IEEEtran}
\bibliography{mybib}

\end{document}